\documentclass[11pt]{article}
\newcommand{\Comment}[1]{{}}
\usepackage{amsfonts,amsmath,epsfig,color,latexsym,graphics}
\usepackage{cite}
\usepackage[textwidth = 430 pt, textheight = 630 pt]{geometry}

\usepackage[active]{srcltx}
\usepackage{graphicx}
\usepackage{bbm}
\usepackage{amssymb}

\Comment{\usepackage{color}
\definecolor{MyDarkBlue}{rgb}{0.15,0.15,0.45}
\usepackage[linktocpage=true]{hyperref}
\hypersetup{
colorlinks=true,
citecolor=MyDarkBlue,
linkcolor=MyDarkBlue,
urlcolor=MyDarkBlue,
pdfauthor={},
pdftitle={
},
pdfsubject={hep-th}
}

\usepackage[numbers,sort&compress]{natbib}
\usepackage{hypernat}}

\def\IZ{\relax\ifmmode\mathchoice
{\hbox{\cmss Z\kern-.4em Z}}{\hbox{\cmss Z\kern-.4em Z}}
{\lower.4pt\hbox{\cmsss Z\kern-.4em Z}}
{\lower1.2pt\hbox{\cmsss Z\kern-.4em Z}}\else{\cmss Z\kern-.4em Z}\fi}
\newcommand{\Z}{\mathsf{Z}\kern -5pt \mathsf{Z}}
\newcommand{\unit}{\mathsf{1}\kern -3pt \mathsf{l}}
\def\one{1\kern -3pt \mathrm{l}}

\def\theequation{\thesection.\arabic{equation}}

\newcommand\ignore[1]{}
\def\one{{\,\hbox{1\kern-.8mm l}}}

\def\Z{\mathbb{Z}}

\newcommand{\Cset}{{\,\,{{{^{_{\pmb{\mid}}}}\kern-.45em{\mathrm C}}}}}

\def\IZ{\mathbb{Z}}
\def\IP{\mathbb{CP}}

\newcommand{\be}{\begin{equation}}
\newcommand{\ba}{\begin{eqnarray}}
\newcommand{\bea}{\begin{eqnarray}}

\newcommand{\ee}{\end{equation}}
\newcommand{\eea}{\end{eqnarray}}
\newcommand{\ea}{\end{eqnarray}}

\newcommand{\nn}{\nonumber}

\def \J {{\cal J}}  

\providecommand{\lsim}{\lesssim}

\begin{document}

\renewcommand{\thefootnote}{\fnsymbol{footnote}}

\makeatletter
\@addtoreset{equation}{section}
\makeatother
\renewcommand{\theequation}{\thesection.\arabic{equation}}

\rightline{}
\rightline{}
   \vspace{0.8truecm}

\vspace{10pt}


\begin{center}
{\LARGE \bf{\sc 
Eliminating ambiguities for quantum corrections to  
\vspace{.1cm}

strings moving in $AdS_4\times \mathbb{CP}^3$
}}
\end{center} 
 \vspace{1truecm}
\thispagestyle{empty} \centerline{
{\large \bf {\sc Cristhiam Lopez-Arcos${}^{a}$}}\footnote{E-mail address: \Comment{\href{mailto:crismalo@ift.unesp.br}}{\tt 
    crismalo@ift.unesp.br}}
{\bf{\sc and}}
{\large \bf {\sc Horatiu Nastase${}^{a}$}}\footnote{E-mail address: \Comment{\href{mailto:nastase@ift.unesp.br}}{\tt 
    nastase@ift.unesp.br}}    }

\vspace{1cm}

\centerline{{\it ${}^a$ 
Instituto de F\'{i}sica Te\'{o}rica, UNESP-Universidade Estadual Paulista}} \centerline{{\it 
R. Dr. Bento T. Ferraz 271, Bl. II, Sao Paulo 01140-070, SP, Brazil}}

\vspace{1truecm}

\thispagestyle{empty}

\centerline{\sc Abstract}

\vspace{.1truecm}

\begin{center}
\begin{minipage}[c]{380pt}{\noindent 
We apply a physical principle, previously used to eliminate ambiguities in quantum corrections to the 2 dimensional kink, to the 
case of spinning strings moving in $AdS_4\times \mathbb{CP}^3$, thought of as another kind of two dimensional soliton. We find that this eliminates
the ambiguities and selects the result compatible with AdS/CFT, providing a solid foundation for one of the previous calculations, which found 
agreement. The method can be applied to other classical string ``solitons''. }
\end{minipage}
\end{center}

\vspace{.5cm}

\setcounter{page}{1}
\setcounter{tocdepth}{2}

\newpage

\renewcommand{\thefootnote}{\arabic{footnote}}
\setcounter{footnote}{0}

\linespread{1.1}
\parskip 4pt

\section{Introduction}

Quantum corrections to solitons have a long and complicated history, and it has proven difficult to find an algorithmic way to calculate them, 
due to regularization-dependent ambiguities. The most studied case, for being the simplest and easiest to analyze, is the kink in two dimensions. 
Studies of its quantum corrections started with \cite{Dashen:1974cj} (see also \cite{Coleman:1974bu,deVega:1976sm}, and supersymmetric 
extensions started with \cite{D'Adda:1977ur,D'Adda:1978mu,Horsley:1978uy,Schonfeld:1979hg})
and still go on (see \cite{Rajaraman:1982is} for basic techniques and references, and \cite{Rebhan:2009nf} for a review of recent results), 
due to the many subtleties present. 
In \cite{Nastase:1998sy} a physical principle was proposed that eliminates the ambiguities and gives a quantum correction consistent (in the 
supersymmetric case) with supersymmetry. 

A seemingly different area that has received a lot of attention lately is quantum corrections to classical (long) strings moving in gravitational 
backgrounds. The reasons for that interest are {\em usually} related to AdS/CFT, since 
one application has been to systems which have a field theory dual admitting a
Bethe ansatz for the dual to the string.\footnote{Of course, systems with Bethe ansatz are interesting in their 
own right, outside the existence of AdS/CFT}
This is useful, since unlike other cases, when we need to invoke supersymmetry to match weak coupling field
theory results to strong coupling gravity results, the Bethe ansatz allows one to have a prediction for the expected quantum correction at strong 
coupling. Then, provided we can trust AdS/CFT and the Bethe ansatz, we have a prediction for the expected quantum correction. 

Of course, the classical string is just a type of solitonic solution in a two dimensional field theory (the sigma model of the string moving in the 
gravitational background), and as such a priori suffers from the same ambiguities as the largely studied kink. From this point of view, one should not 
be surprised that early calculations for the corrections to a spinning string in $AdS_4\times \mathbb{CP}^3$ gave different results, and 
apparently incompatible with AdS/CFT  \cite{McLoughlin:2008ms,Alday:2008ut,Krishnan:2008zs,Gromov:2008fy}.
In \cite{McLoughlin:2008he}, a calculation was proposed that matches with AdS/CFT expectations.\footnote{Later proposals were made on how it could be possible to match 
with AdS/CFT other calculations as well.} See 
\cite{Zarembo:2009au,Bandres:2009kw,Mikhaylov:2010ib,Abbott:2010yb,Astolfi:2011ju,Abbott:2011xp,Astolfi:2011bg,Beccaria:2012qd} for later 
related works. 

However, the calculation in \cite{McLoughlin:2008he} still suffers from the same {\em a priori} ambiguities, and it amounts to a particular choice of
regularization for them, whose only justification is a posteriori, through matching with AdS/CFT. It is the purpose of this paper to provide a 
justification for that calculation, by taking the physical principle of \cite{Nastase:1998sy} and applying it to classical strings. We will show 
that its use for the model in \cite{McLoughlin:2008he} eliminates the ambiguities implicitly hidden there, and thus offers the possibility of extending 
the same method to other classical string solutions. 

The paper is organized as follows. In section 2 we explain our method based on a physical principle, in section 3 we review how it applies to the 
case of the two dimensional kink, in section 4 we apply it to the spinning string in $AdS_4\times \mathbb{CP}^3$, and in section 5 we conclude.

\section{The method}

As is well known, one-loop corrections to the energy of the vacuum are equivalent (via the exponentiation of one-loop determinants)
 to sums of zero-point fluctuations, for a bosonic mode
$\sum \frac{1}{2}\hbar \omega^B$. If we have fermionic modes, they will contribute with $-\sum \frac{1}{2}\hbar \omega^F$.
These fluctuations give rise in particular to the Casimir energy, which is the difference in this zero-point energy between the infinite space
and a space of finite size. Of course, this answer is a priori ambiguous ($\infty -\infty$), and moreover highly divergent. Generically, 
$\omega\sim n$ implies a quadratic ($n^2$) divergence. 

The same idea applies when we calculate the quantum mass of some soliton, generically denoted $\phi_{sol}(x)$, with classical mass $M$. 
We have to calculate the 
fluctuations in the presence of the soliton, i.e. eigenmodes around $\phi_{sol}(x)$, and subtract the fluctuations in its absence (in the vacuum). 
An extra factor to take into account is renormalization. In terms of Feynman diagrams, we know we have counterterms, which correspond to 
renormalizing the parameters of the theory, for instance a bare mass parameter $m_0$ becomes the renormalized mass $m$. When going to the 
fluctuation representation, a useful way of encoding the counterterm for the energy, $\delta M$, is by the variation of the classical mass $M$ 
when expressed in terms of unrenormalized parameters like $m_0$, vs. renormalized ones like $m$, with a result linear in $m_0-m=\delta m$, 
where for $\delta m$  we need to take the result of the one-loop Feynman diagramatic calculation.

Therefore generically the one-loop contribution to the quantum mass of a soliton is given by
\be
E_1=\frac{1}{2}\sum_n(\omega_n^B -\omega_n^F)-\frac{1}{2}\sum(\omega_n^{(0)B} -\tilde{\omega}_n^{(0)F})+\delta M.
\label{e_1loop_gen}
\ee
where the $\omega_n^B, \omega_n^F$ are the frequencies coming from the bosonic and fermionic parts of the action, respectively, labelled by an 
integer $n$, and the $(0)$ refers to the vacuum, i.e. without the soliton solution.  

This expression contains ambiguities. 
The first type of ambiguities is due to the fact that we have
generically the $\infty -\infty$ difference of quadratic divergences (if at large $n$, $\omega_n \sim n$, then $\sum 
\omega_n\sim n^2$), which a priori will be linearly divergent ambiguities, not even constant ambiguities. Here we should note that we would be 
tempted to say that if we have something like, say, $M=\sum_n\sqrt{1+m^2/n^2}-\sum_n 1$, this is the same as $M=\sum_n m^2/(2n^2)$ which is finite. 
But this in fact amounts to a particular choice of regularization scheme. One needs more information to be precise about which regularization it
is, but this would basically be part of the 
mode number regularization, if we would have instead of $n$, a $k_n$ together with a relation between $n$ and $k$. Mode number regularization 
means that we identify each mode in a sum with another mode in the other sum, effectively giving the summation operator as a common factor.  
In general however, we are not allowed to make the $\sum_n$ common if both sums are infinite. In terms of choosing a cut-off, there are always at least 
two ways to regularize, mode number cut-off (which corresponds to making the {\em sum} common) and energy/momentum cut-off. The second, 
choosing the same upper {\em energy} instead for the two sums
(convert sum over $k_n$ to integrals over $k$ and identify the variable $k$ in the two sums) gives different results if the sums are infinite
\cite{Rebhan:1997iv}. Note that even in usual quantum field theory divergent integrals we can have this situation, just that usually one 
doesn't think about it. For instance, if $\int f$ and $\int g$ are UV divergent, then $\int^\Lambda(f-g) $ automatically means that we take the same
cut-off $\Lambda$ for $f$ and $g$, but we could in principle choose $\int^\Lambda f-\int^{\Lambda +a} g$, giving a different result. There might be 
situations where this is necessary. 

Yet another type of ambiguity is related to the existence of different types of possible boundary conditions, in turn determining different functions 
$k_n$, or $k(n)$ in the continuum limit.

But we want a physically unambiguous way to determine the correct regularization and boundary conditions. In \cite{Nastase:1998sy} a physical 
principle was 
used to fix both. The principle can be simply formulated by saying that the non-trivial topology of the soliton boundary does not introduce any extra 
energy. 

The first part of the method 
involves the notion of topological boundary condition, i.e. that the boundary condition should not introduce boundary-localized
energy (surface effects), thus fixing one type of ambiguity. 
For scalar fields, the boundary conditions should be compatible with the classical solution (if the classical solution is antiperiodic, then 
so must the boundary condition for fluctuations), but for fermions and higher spins we need to be more careful. 
The method was described in detail in \cite{Nastase:1998sy}: consider the symmetries of the action and the symmetries of the solution. 
For the kink, the action has a $\{\phi\rightarrow -\phi, \psi\rightarrow\gamma_3\psi\}$ symmetry and a $\psi\rightarrow -\psi$ symmetry, and 
the kink solution is antisymmetric in $\phi$. Then e.g., the fluctuations around the kink solution have 
\be
\phi(-L/2)=-\phi(L/2);\;\;\;
\phi'(-L/2)=-\phi'(L/2);\;\;\;
\psi(-L/2)=(-1)^q \gamma_3\psi(L/2)
\ee

The second part is that when we take the classical soliton mass to zero, specifically by taking a relevant mass scale on which it depends, 
like the $m$ above, to zero, the quantum mass of the soliton should also go to zero, such that there is no mass depending purely on topology, i.e. 
localized at the boundary. That in turn means that we can calculate instead of the soliton mass, its derivative with respect to the relevant mass scale 
$m$, thus reducing the UV divergence of the result, and obtaining a ``derivative regularization''. For instance, in the example above, 
$\partial M/\partial m=\sum_n m/(n^2\sqrt{1+m^2/n^2})$ is now indeed finite and unambigous.

We should emphasize that it is not guaranteed that this procedure eliminates ambiguities in general, since taking only one derivative may not 
reduce the divergence sufficiently. Nevertheless, we hope that in many cases of interest, the result is unambiguous. Note that taking more derivatives
will in general reduce further the divergence, but it is not clear if there is a physical principle that will correspond to this modified 
prescription.

We can therefore define our procedure as follows: Find the soliton solution $\phi_{sol}(x)$, find the frequencies of fluctuations around it, and the 
renormalization of the relevant mass parameters. Then find the relevant mass parameter to define derivative regularization with respect to it, and 
topological boundary conditions. Ideally, the resulting quantum mass should be well-defined and unambigous, and we should be able to calculate it.
We will see that in the simple $\lambda\phi^4$ kink case it is indeed true, however the string soliton case is more complicated. We can prove that 
the resulting answer is unambiguous, but we will still need to employ the same procedure used in \cite{McLoughlin:2008he} to calculate it.

\section{One-loop mass for the kink in $\phi^4$ theory}

We want first to understand how this method applies to the kink solution of the $\phi^4$ theory in two dimensions. Here we review \cite{Nastase:1998sy}.

The theory has the Lagrangian
\be
{\cal L}=-\frac{1}{2}(\partial_{\mu}\phi)^2-\frac{\lambda}{4}(\phi^2 -\mu^{2}_{0}/\lambda)^2.
\ee

There are two degenerate vacuum states (trivial solutions), $\phi=\pm \mu_{0}/\sqrt{\lambda}$, and therefore 
topologically nontrivial, localized solutions 
(kinks) must tend to $\pm \mu_{0}/\sqrt{\lambda}$ as $x\longrightarrow \pm\infty$.

We thus have two nontrivial, stable, finite energy solutions, the kink and anti-kink
\be
\phi_{K,\bar{K}}=\pm\mu_{0}/\sqrt{\lambda}\tanh[\mu_{0}(x-x_0)/\sqrt{2}],
\ee
with classical mass $M_0=2\sqrt{2}\mu_{0}^{3}/3\lambda$.

The eigenfrequencies of small fluctuations around the vacuum (trivial sector) are
\be
\tilde{\omega}_n=\sqrt{\tilde{k}_{n}^{2}+m^2},
\ee
and the allowed values for $k_n$ come from the condition $k_n L = 2\pi n$, where $L$ is the size of the one dimensional spatial box in which we put the 
system.

The eigenfrequencies for small fluctuations around the kink (nontrivial sector)
have the same expression as the trivial vacuum ($\omega_n=\sqrt{k_{n}^{2}+m^2}$), but  
the condition for the allowed values of $k_n$ has a different form
\be  
k_n L+\delta(k_n)=2\pi n,\label{allowed}
\ee
where the explicit form of the phase shifts $\delta(k)$ is
\be
\delta(k)=\left(2\pi-\arctan\left(\frac{3m\mid k\mid}{m^2-2k^2}\right)\right)\epsilon(k),
\ee
and is obtained from the explicit scattering solutions in the potential generated by the perturbation around the kink.

In the case of fermionic fluctuations (for a supersymmetric version of the kink), which are 2 component vectors, 
there is a further phase shift $\theta(k)$ giving a $e^{\pm i\theta(k)/2}$
relative factor at $\pm \infty$ between the two components.

As we mentioned in the previous section, the one-loop counterterm for the soliton mass $M$ comes from varying the classical $M$ under the renormalization $\delta m=m_0-m$, 
and gives 
\be
\delta M=\frac{3m}{4\pi}\int\frac{dk}{(k^2+m^2)^{1/2}} \ ,\ \ \ \ m^2=2\mu^2.
\ee

Now we have the necessary ingredients for the calculation of the one-loop correction to the energy. Substituting the frequencies in
the expression for the 1-loop correction
\be
E_1=\frac{1}{2}\sum\omega-\frac{1}{2}\sum\tilde{\omega} +\delta M,
\ee
we obtain
\be
E_1=\frac{1}{2}\sum\sqrt{k_{n}^{2}+m^2} -\frac{1}{2}\sum\sqrt{\tilde{k}_{n}^{2}+m^2}+\delta M.
\ee
As we can see, even after the subtraction, the sum is linearly divergent. To apply our derivative regularization, we must find
the mass parameter which takes the soliton mass to zero, and take a derivative with respect to it. In this case, it is obvious, namely the 
mass parameter is $m$.  We then differentiate the energy with respect to $m$, perform the summation and integrate back with respect to $m$.

That will get rid of both linearly and logarithmically divergent ambiguities. The physical principle then dictates that the constant of integration 
is zero.

Taking the derivative, we obtain
\be
\frac{dE_1}{dm}=\frac{1}{2}\sum\frac{d\omega}{dm}-\frac{1}{2}\sum\frac{d\tilde{\omega}}{dm} +\frac{d\delta M}{dm},
\ee
where 
\bea
\frac{d\tilde{\omega}}{dm}&=&\frac{m}{\sqrt{\tilde{k}^{2}_n+m^2}},\cr
\frac{d\omega}{dm}&=&\frac{1}{\sqrt{k^{2}_{n}+m^2}}\left(m+\frac{k^{2}_{n}}{Lm}\delta'(k_{n})\right).
\eea
The sums are less divergent now, and can be turned into integrals by taking into account the conditions for the allowed $k_n$s (\ref{allowed}),
obtaining a finite result. Integrating back with respect to $m$ we will get a constant of integration, but applying the physical 
principle, the constant is zero. Therefore finally the 1-loop energy correction is
\be
E_1=m\left(\frac{1}{4\sqrt{3}}-\frac{3}{2\pi}\right).
\ee

\section{One-loop corrections to spinning strings on AdS$_4\times \IP^3$}

\subsection{Applying the method}

We now try to apply the same method to the classical (long) string on the background AdS$_4\times \IP^3$. 
This can be thought of just as another 2d field theory, with action
\be
S = \frac{R^2_{\rm AdS}}{4 \pi}\int \ d\tau \int_0^{2\pi}
d\sigma\ \sqrt{g}g^{ab } \left(
      G_{\mu\nu}^{\rm AdS}\partial_a X^{\mu} \partial_b X^{\nu}
   +4 G_{\mu\nu}^{\IP^3}\partial_a X^{\mu} \partial_b X^{\nu}
                         \right)~.
\label{bose_action}
\ee
As we have discussed, the first step is to understand the 2d vacuum and soliton solution.
The (trivial) ``vacuum'' corresponds to the point-like string, equivalent to $\phi=\phi_0$=constant for the $\lambda \phi^4$ model.
The nontrivial soliton whose mass we want to calculate is a spinning string solution, with nontrivial $X^\mu(\sigma,\tau)$. 

The computation of the 1-loop energy correction for this soliton was done in different ways, obtaining different results. The calculation of 
\cite{McLoughlin:2008he} 
gave the correct result matching the expectation from AdS/CFT, but there was no {\em a priori} reason why it should be correct, given the 
implicit choice of regularization scheme needed to obtain it. We will therefore try to identify the ambiguities as above.

In order to apply our method, we note several complications with respect to the kink case. We note that (\ref{bose_action}) is a 
non-linear sigma model, and we have no potential, so formally it looks different from the kink. That however means we can avoid at least one 
ambiguity from the kink case. No potential means that the phase shifts $\delta(k)$ and $\theta(k)$ are not present, so at least the ambiguity of 
boundary conditions (related to non-zero $\delta(k)$ and $\theta(k)$) is not there. It is also lucky, since for the calculation of $\delta(k)$ 
and $\theta(k)$ we would need the full solutions, which as we will see are hard to find.
The only ambiguity we still have is the UV divergence.

To deal with that, we need to define our physical regularization. But instead of the mass parameter $m$ of the kink,
we will have several parameters (which come in the solution), and we have to carefully analyze which can be varied in order to relate the energies and use the derivative regularization. 
Note however that there are no parameters {\em in the action} (\ref{bose_action}) (other than $R_{AdS}$ which multiplies the whole action, 
so is not relevant), so the parameters of relevance will just characterize the vacuum solution. That also means that there are no counterterm 
contributions, since the only possible counterterm could be for $R_{AdS}$, which is not renormalized.

The parameter we want needs to be something that when equal to zero, takes the classical mass of the long string to zero, 
but also something that, like $m$ for the kink, is normally non-zero in the vacuum. 

An extra complication will be, as we will see, that it is not possible to find the full solutions for the eigenfrequencies, only as an  
expansion in a large parameter $\omega$. But then it matters how $n$ is related to $\omega$; in particular, the expansion is not valid for 
$n>\omega$, which corresponds to the UV divergence we want to analyze. So the only goal we will have is to show that the physical derivative 
regularization obtained as above selects the regularization implicit in \cite{McLoughlin:2008he}. 
In order to actually compute the quantum correction, we will still need 
to use the same procedure as in \cite{McLoughlin:2008he}.

In the following sections we will perform first the classical analysis of the model, then we will find the frequencies, and 
finally apply the derivative regularization.

\subsection{The nontrivial soliton}

In this subsection we will analyze the spinning string in AdS$_4\times \IP^3$. We will see that there are several parameters present in 
this nontrivial solution, but there are relations between them due to the Virasoro constraints, so our search for the parameter that 
is nonzero in the vacuum, but takes the soliton mass to zero when it equals zero (the analog of the mass parameter $m$ for the kink), 
will be highly constrained. 
The conserved quantities, like the energy, which here has the meaning of ``soliton mass'' modulo an additive constant, 
will be dependent on these parameters. 
An important technical detail is that the Virasoro constraints are complicated, so we can only solve them perturbatively in certain limits, 
hence the same will happen for the energy (``soliton mass'').

The classical analysis for the string in this background have been done completely (see \cite{McLoughlin:2008he} or \cite{McLoughlin:2008ms}), so here we will review the main points. The bosonic part of the action for the spinning string is the one in (\ref{bose_action}), which can be split as
\be
S=S_{\rm AdS_4}+S_{\IP^3},
\ee
and the background metrics appearing in the nonlinear sigma model are
\bea
\label{ads_metric}
ds^2_{{\rm AdS}_4}&=&
-\cosh^2\rho\ dt^2+d\rho^2+\sinh^2\rho 
\left(d\theta^2+\sin^2\theta d\phi^2\right)\ , 
\\
\label{p3_metric}
ds^2_{\IP^{3}}&=& d\zeta_1^2
+\sin^2\zeta_1\left[ d\zeta_2^2+\cos^2\zeta_1
\left(d\tau_1+\sin^2\zeta_2\left(d\tau_2+\sin^2\zeta_3d\tau_3\right)\right)^2
\right. \nn\\
& &\left. 
+ \sin^2\zeta_2\left( d\zeta_3^2+\cos^2\zeta_2
\left(d\tau_2+\sin^2\zeta_3d\tau_3\right)^2
+\sin^2\zeta_3\cos^2 \zeta_3 d\tau_3^2 \right)\right]~~.
\eea

Here we have factored out the scale of the metric
\be
R^2_{\rm AdS}=\sqrt{\bar \lambda}=\sqrt{2\pi^2\lambda}=\sqrt{2\pi^2\frac{N}{k_{\rm CS}}}
\ee
which is very large (very large $\bar \lambda$, though finite).

The soliton we are interested in was found in \cite{Park:2005ji}. It is a rotating string lying in an AdS$_3\times$S$^1$ subspace of 
AdS$_4 \times \IP^3$, which from the point of view of the 2d worldsheet looks like a soliton with 
\bea
&&\bar{t}=\kappa\tau , \ \ \bar{\rho}=\rho_* , \ \ \bar{\theta}=\frac{\pi}{2}
, \ \ \bar{\phi}=w\tau+k\sigma , \cr
&&\bar{\tau_1}=\bar{\tau_3}=\frac{1}{2}(\omega\tau + m\sigma), \ \  
\bar{\tau_2}=0 , \cr
&&\bar{\zeta_1}=\frac{\pi}{4} , \ \ \bar{\zeta_2}=\frac{\pi}{2}, \ \  
\bar{\zeta_3}=\frac{\pi}{2} .
\label{sol}
\eea

Unlike the kink case or usual quantum field theory, now we have also gravity on the worldsheet, which in the conformal gauge manifests 
itself in the presence of the Virasoro constraints $T_{ab}=0$.
For the solution (\ref{sol}), we have an equation of motion 
\be
w^2-\kappa^2 -k^2=0,
\ee
and the Virasoro constraints reduce to
\bea
r_1^2wk+\omega m&=&0,\cr
-r_0^2\kappa^2+r_1^2(w^2+k^2)+\omega^2 m^2&=&0.
\eea 
They can be solved perturbatively, as done in \cite{McLoughlin:2008he}, in a certain limit that we will define shortly.

The charge densities are
\be 
{\cal E}=\int^{2\pi}_0  \frac{d\sigma}{2\pi}\ r_0^2  \kappa 
        ={\rm r}_0^2\kappa \   , \quad 
{\cal S}=\int^{2\pi}_{0} \frac{d\sigma}{2\pi}\ {r_1^2}{\rm w}
        = {\rm r}_1^2 {\rm w} \ ,\qquad 
{\cal J}_2={\cal J}_3=\int^{2\pi}_{0} \frac{d\sigma}{2\pi} \  \omega
        =\omega,
\label{charges}
\ee
so that  the classical energy, spin and 
the charges under the second and third Cartan generators 
of $SO(6)$ are 
\be
E_0 =\sqrt{\bar\lambda}\, r_0^2 \kappa\ , 
\qquad  \quad  
S=\sqrt{\bar\lambda}\,{\rm r}_1^2 {\rm w}\ , 
\qquad \quad 
J\equiv J_2=J_3=\sqrt{\bar\lambda}\,\omega~ ,
\ee
where $r_0=\cosh\rho_*$.

The limit we use to solve the constraints (following \cite{McLoughlin:2008he})
and find some relations between the constants consists in taking large spin ${\cal S} $ and large angular momentum $ {\cal J}$,
with their ratio $u$ (and also $k$) held fixed, i.e.
\be
{\cal S},\ \,{\cal J}\rightarrow\infty\ , \ \ \ \ \ \ \ \ \ \ \
u=-\frac{m}{k}=\frac{\cal S}{ {\cal J}}=\frac{S}{ J}={\rm fixed} \ .
\label{scaling}
\ee

For this solution, the expansion of the classical energy at large $\J=\omega$ and thus large angular momentum
$J=\sqrt{{\vphantom{{{}^{|^A}}}}\bar\lambda} {\cal
J}=\sqrt{{\vphantom{{{}^{|^A}}}}\bar\lambda}\, \omega $ is given by
\bea
E_0&=&  S +  J 
    + \frac{ {\bar\lambda} }{2J} k^2u(1+u) 
    -  \frac{{\bar \lambda}^2}{8J^3}k^4 u(1+u)(1+3u+u^2)\nn\\
& & +\ \frac{{\bar \lambda}^3}{16J^5}k^6 u(1+u) (1+7u+13u^2+7u^3+u^4)
+{\cal O}\left(\frac{1}{J^7}\right)\ . \nn\\
\label{E0}
\eea

We will see later that this large $\omega$ limit is also needed to have a workable form for the eigenfrequencies around the classical solution.

On top of this limit, in the next subsection we will use another perturbative expansion which will have as a limit a trivial sector (``vacuum'').
We will later see that we need to be only a bit away from this new limit (i.e., to be in the perturbative expansion) in order to be able to use our 
regularization procedure.

\subsection{The vacuum solution}

Since the two dimensional soliton we are interested in corresponds in spacetime to a long spinning string, it follows easily that the trivial solution 
(``vacuum'') has to be a point-like string.  Guided by the BMN limit \cite{Berenstein:2002jq}, where we also have perturbations around a state with 
large $J$, we know that the eigenvalues of the Hamiltonian, corresponding to perturbations 
around a BPS state, are the equivalent of the soliton mass, and therefore we look for states of lowest $E-J$ as the vacuum. We then  
vary the parameters in the nontrivial solution to obtain such a vacuum. 
This Hamiltonian is, as we saw, $E_0-J$ from (\ref{E0}),
where $S\sim r_1$ and $u\sim m$. The smallest value is then obtained for
\bea
r_1 , m &\longrightarrow & 0, \nn\\
r_0 &\longrightarrow &1. \nn \\  
\label{valim}
\eea
which implies in particular very small $S$ as well (relative, since we formally took $S\rightarrow \infty$ before, though note that $\bar\lambda$ is 
large in $S=\sqrt{\bar\lambda}r_1^2w$), with everything else ($J,\omega,k,\kappa$) kept fixed in this second limit.

Then we obtain the ``soliton mass'' in the vacuum $E-J=0$, as we wanted.

Taking these limits directly on the spinning solution we indeed get then the point-like string, the trivial solution we were looking for. 
Now that we have both solutions we can proceed to analyze quantum fluctuations around them.

\subsection{The spectrum of quadratic fluctuations}

\subsubsection{Bosons}

To find the characteristics frequencies we expand the action (\ref{bose_action}) around the solution (\ref{sol}). For the bosonic 
fluctuations we have six scalars corresponding to motion on $\IP^3$: one is massless, other four  degrees of freedom 
give the same result,
\be
p_0=\sqrt{p_1^2+\frac{1}{4}(\omega^2- m^2)}\ , 
\ee
and the last one gives
\be
p_0=\sqrt{p_1^2+(\omega^2- m^2)}~.
\ee
From the scalars corresponding to motion in AdS space we find one massless degree of freedom, one massive one with
\be
 p_0=\sqrt{p_1^2+\kappa^2}\ , 
\ee
and two fluctuations whose dispersion relation is given by the roots
of the quartic equation
\be
(p_0^2-p_1^2)^2+ 4 r_1^2 \kappa^2 p_0^2
-4\left(1+r_1^2\right)\left(\sqrt{\kappa^2+k^2}\ p_0-k p_1\right)^2=0\ .
\label{boo}
\ee

We can find the explicit solutions to this equation (though they do not give much information), but only when we expand in large $\omega$.

\subsubsection{Fermions}

For the fermionic part the spectrum contains four different frequencies, each
being doubly-degenerate. Two such pairs have frequencies
\be
(p_0)_{\pm 12}=\pm \frac{{\rm r}_0^2 k\kappa m}{2(m^2+{\rm r}_1^2k^2)
}
+\sqrt{(p_1\pm b)^2+(\omega^2+k^2{\rm r}_1^2)}\ , ~~~~~~~~~
b\equiv-\frac{\kappa m}{{\rm w}}\frac{{\rm w}^2-\omega^2}{2(m^2+{\rm
r}_1^2k^2)
},
\ee
while the frequencies of the other two pairs are solutions of the
equation
\be
(p_0^2-p_1^2)^2+  {\rm r}_1^2 \kappa^2  p_0^2
-\left(1+{\rm r}_1^2\right)\left(\sqrt{\kappa^2+k^2}\ p_0-k p_1\right)^2=0~~.
\label{kpq}
\ee
which can be solved in the same limit as in the bosonic case. 

With the bosonic and fermionic frequencies we can start to calculate the quantum corrections, formally defined as in (\ref{e_1loop_gen}).
But in order to do that, we must apply a regularization technique, specifically the derivative regularization previously defined. For that, 
we need to find the parameter that plays the role of $m$ for us.

\subsection{Physical limit and regularization}

We want the parameter to lead to $E=J$ as it goes to zero, but be otherwise finite in the vacuum. Since
\bea
E_0&=&  S +  J 
    + \frac{ {\bar\lambda} }{2J} k^2u(1+u) 
    -  \frac{{\bar \lambda}^2}{8J^3}k^4 u(1+u)(1+3u+u^2)\nn\\
& & +\ \frac{{\bar \lambda}^3}{16J^5}k^6 u(1+u) (1+7u+13u^2+7u^3+u^4)
+{\cal O}\left(\frac{1}{J^7}\right)\ . \nn\\
\eea
we could try $u$ or $k$ only, as we have $S=Ju$. Note one subtlety here: we have $E_0=E_0(S,J,\bar\lambda, u,k)$, however $u=S/J$ so 
there is an ambiguity in the split of $E_0$ (how to we isolate the $S$ dependence, when we could always write any $u$ as $S/J$). We can consider that 
the $S$ term is the one that is independent on $k$, which will be useful shortly.

However, $u$ is not a good parameter, since it becomes always zero in the vacuum. 
On the other hand, $k$ stays fixed in the vacuum, yet $k\rightarrow 0$ keeping everything else ($u,J, \omega, \kappa$) fixed
gives $E_0\rightarrow S+J$. That is then not enough, and we need to supplement our original definition of the nontrivial vacuum with 
$u$ small, and therefore also $w,S,m$ small, i.e. in the perturbative expansion away from the vacuum.

Therefore $k$ is the parameter that relates the two energies. One more subtlety to note is that, since we will use the large $\omega$ expansion, 
and since 
\be
u=\frac{S}{J}=\frac{r_1^2w}{\omega}\lsim \frac{1}{\omega},
\ee
by doing the $1/\omega$ expansion first, we will not be able to match terms linear in $u$, as we will explain better later.

We should note that it was crucial that there were at least two parameters, $J$ and $k$: $J$ to guarantee a long string, with large $J$ giving 
a perturbation theory, and $k$ to differentiate with respect to it. It is our hope that this is more general for long strings, with something 
like $J$ guaranteeing a long string, and something like $k$ giving the ``shape", allowing us to differentiate with respect to it.

We are finally ready for the calculation of the quantum correction to the energy.

\subsection{Quantum correction to the energy}

The one-loop energy correction was thought of as \cite{Frolov:2002av} 
\be
E_1=\frac{1}{\kappa}\langle\Psi\vert H_2\vert\Psi\rangle,
\ee
where $H_2$ is the Hamiltonian for the quadratic fluctuations, but subtleties arose that were not well appreciated.

In order to understand what the issues are, we first review a few facts about previous calculations. 

First, previous calculations have not taken into account the  trivial sector or ``vacuum'' (cf. (\ref{e_1loop_gen})), but considered only 
$E_1=\frac{1}{2}\sum_n(\omega_n^B -\omega_n^F)$. Of course, at the classical level that does not matter, but it does matter at the quantum 
one-loop level. As we will see, removing the contribution of the trivial sector from the sum will help to the cancellation of some ambiguities. 

Second, since one gets divergent sums in $E_1$, a regularization scheme is necessary, and various calculations gave regularization-dependent 
results \cite{McLoughlin:2008ms,Alday:2008ut,Krishnan:2008zs,Gromov:2008fy,McLoughlin:2008he,Bandres:2009kw}. 
In the calculations of \cite{McLoughlin:2008ms,Alday:2008ut,Krishnan:2008zs} the sum was turned into an integral, after which a cut-off was 
introduced and the integral sign given as a common factor, effectively choosing a form of energy/momentum cut-off regularization, as we 
explained in section 2. In \cite{Gromov:2008fy} a different regularization was chosen, where one combines a mode number cut-off with a certain 
grouping of terms: instead of $\sum_n(\omega_n^{Bose}-\omega_n^{Fermi})$, one forms combinations called $\omega_n^{heavy}$ and $\omega_n^{light}$
and then a certain $n-$dependent combination of $\omega_n^{light}$ is added to $\omega_n^{heavy}$, and the resulting sum over $n$ is turned into 
an integral. This regularization gave a different result from the previous one. More recently, in \cite{Bandres:2009kw}, a modification of 
the regularization in \cite{Gromov:2008fy} was given, with different combinations of $\omega_n^{heavy}$ and $\omega_n^{light}$. 

Yet another type of regularization was considered in \cite{McLoughlin:2008he}, where a regularization 
method used successfully in the case AdS$_5\times S^5$ \cite{Beisert:2005cw} was applied, together with a physically motivated redefinition of 
the coupling constant. 
The result of \cite{McLoughlin:2008he} is in agreement with AdS/CFT, so it was considered correct, but {\em 
a priori} we did not know which regularization scheme to choose to obtain an unambigous result, since as we saw different schemes can lead to 
different results, exactly as in the case of the 2d kink. We can use matching with AdS/CFT only as a kind of a posteriori check, exactly as one 
used the saturation of the BPS bound for the 2d supersymmetric kink (where both the mass and the central charge of the kink get renormalized 
in the same way).

In what follows we take a large $\omega$ expansion for both trivial and nontrivial sectors, and 
we will focus on the leading order in this expansion. As mentioned above, this will force us to take a small $u$ expansion as well, and we 
can only say something about the leading term in the $u$ expansion. 

More importantly, in \cite{McLoughlin:2008he} 
it was explained that if we expand in $1/\omega$, since we can allow any value of $p_1=n$, we have two regions 
for the expansion: a) $1/\omega\rightarrow 0$ with $n$ fixed, i.e. $n\ll \omega$, for which we still have a discrete sum; and b) $n,\omega\rightarrow
\infty$, with $x=n/\omega$=fixed, for which we can replace the sum with an integral. It was then noticed that while both regions contain divergences, 
the divergence of one can be identified with the divergence of the other, and can be dropped, obtaining a finite result. What we want to show here 
is that the ambiguity inherent in this procedure is removed by our method. 

What we would have liked to do is take first the derivative with respect to $k$, and then do the sum over $n$, maybe with the same $1/\omega$ 
expansion, but this turns out to be prohibitively difficult, so we will be forced to follow the same analysis as \cite{McLoughlin:2008he} 
once we prove that our method eliminates the ambiguities.

We will start by analyzing region a), where we have discrete sums, and where 
\be
u\lsim \frac{1}{\omega}\ll \frac{1}{n}.
\ee

The trivial sector (``vacuum'') is simpler, and illustrates the point well, so we will start with it.
The sum of bosonic frequencies (bosonic summand) in the trivial sector is 
\be
\sqrt{w^2-n(2k-n)}+\sqrt{n(2k+n)+w^2}+4 \sqrt{n^2+\frac{\kappa^2}{4}}+2\sqrt{n^2+\kappa^2}.
\ee
Replacing the perturbative solutions of the Virasoro constraints and expanding in $\omega$, we get
\be
6\omega +\frac{6n^2+k^2(u(u+2)-1)}{\omega}.
\ee
A similar procedure for the fermionic summand (minus the sum of fermionic frequencies) gives
\be
-6\omega -\frac{12n^2+k^2(u+1)^2 (u(u+2)-1)}{2\omega}.
\ee
Taking the sum of the two expressions to obtain the summand, we get terms like $n^2-n^2$ and $\omega -\omega$ (since $\omega>n$, these are of the same
type), which are ambiguous, but they will 
be cancelled after taking the derivative with respect to $k$. 
After the derivative with respect to $k$, the trivial sector summand $\tilde{e}(n)$ gives
\be
\frac{\partial \tilde e(n)}{\partial k}\equiv \tilde{e}_k(n)=-\frac{k(1-u(2+u))^2}{\omega}+{\cal O}\left(\frac{1}{\omega^3}\right).
\label{tsk_1_leading}
\ee
with no $n^2-n^2$ and $\omega-\omega$ ambiguities.

Moving on to the nontrivial sector, the leading terms in the large $\omega$ expansion of the nontrivial sector summand are
\bea
&&e(n)=\frac{1}{2 \omega}
\bigg[ n\bigg(3n-4\sqrt{n^2+k^2u(1+u)}+\ \sqrt{n^2+4k^2u(1+u)}\ \bigg) \nn \\
&& \ \ \ \ \ \ \ \ \ \ \ \ \ \ \ \ \ \ \ \ \ \ \ \
- \ k^2(1+u)(1+3 u)\bigg]+{\cal O}\left(\frac{1}{\omega^3}\right),
\label{nts_1_leading}
\eea
and it can be seen that again terms like $n^2-n^2$ appear, but they are again cancelled by taking the derivative with respect to $k$. 
After $\partial/\partial k$, the summand of the nontrivial sector gives
\be
\frac{\partial e(n)}{\partial k}\equiv e_k(n)=-\frac{2k(1+u)(1+3 u) -\left( \frac{4knu(1+u)}{\sqrt{n^2+k^2u(1+u)}}-\frac{4knu(1+u)}{\sqrt{n^2+4k^2u(1+u)}}\right)}{2 \omega}+{\cal O}\left(\frac{1}{\omega^3}\right). 
\ee
We note that even at $u=0$, there is a constant piece that would give a divergence when summed over $n$, however it is the same one as in 
the trivial sector summand (\ref{tsk_1_leading}), so by subtracting the two we get rid of the last potential ambiguity. 

We finally get
\be
e_k(n)-\tilde e_k(n)=\frac{1}{\omega}\left( ku(u(u(4+u)-1)-8)+\frac{2knu(1+u)}{\sqrt{n^2+k^2u(1+u)}}-\frac{2knu(1+u)}{\sqrt{n^2+4k^2u(1+u)}}\right).
\label{sumsummand}
\ee
It would seem that we still have a divergence after we take the sum, but we need to remember that $u\ll 1/n$, so these terms linear in $u$ do 
not give rise to divergences in this limit (or another way of saying it is that they belong to the omitted higher order terms in $1/\omega<1/n$). 

The final result for the one-loop correction to the energy coming from region a) is the sum over (\ref{sumsummand}), integrated over $k$ (with 
zero constant of integration). 

There is a certain subtlety here, since in the end we want to calculate a correction to the energy that will turn out to have contributions 
linear in $u$, 
but as we mentioned, our only purpose (given our technical, i.e. calculational, limitations) is to show that the procedure of \cite{McLoughlin:2008he}
becomes unambigous if we consider our physical principle. 

Let us now analyze the result of \cite{McLoughlin:2008he} and compare to what we get. 
Expanding (\ref{nts_1_leading}), now called $e^{sum}(n)$ to emphasize that we are
in region a), at large $n$ we get
\be
e^{sum}(n)=\frac{1}{2\omega}\left(-k^2(1+u)(1+3u)-\frac{3}{2n^2}k^4u^2(1+u)^2+...\right)+{\cal O}\left(\frac{1}{\omega^3}\right),\label{sumprev}
\ee
where the first term becomes divergent when summed over $n$ (singular piece) and the second term becomes regular. The divergence and hidden 
ambiguities implicit in (\ref{sumprev}) were eliminated in our result (\ref{sumsummand}).

On the other hand, in region b), with $\omega/n=x$=fixed, the expansion of the summand, now denoted $e^{int}(x)$ gives \cite{McLoughlin:2008he}: 
\bea
\label{eqn:integrand_freq_x}
e^{\rm int}(x)&=& \frac{k^2(1+u)}{2 \omega} \left(\frac{1+u(3+2 x^2)}{(1+x^2)^{3/2}}
- 2\frac{1+u(3+8 x^2)}{(1+ 4 x^2)^{3/2}}\right)\nn\\
& &\kern-20pt- \ \frac{k^4(1+u)}{32\omega^3 x^2} \Big[\frac{1}{(1+ x^2)^{7/2}}
\left(32u^2(1+u)+(7+u(77+u(221+135u)))x^2\right.
\nn\\& &\left.+4(-7+u(-7+u(29+21u)))x^4+16u(1+u(3+u))x^6+16 u(1+u)x^8\right)\nn\\
& & -\frac{8}{(1+4x^2)^{7/2}}\left(u^2(1+u)+(1+3u(5+u(11+5u)))x^2\right.\nn\\
& &\left.+8(-1+3u)(2+u(4+u))x^4+64u(2+3u)x^6+256u(1+u)x^8\right)\Big]+{\cal O}\left(\frac{1}{\omega^3}\right).\nn\\
\eea
Note that in computing this expression we have also assumed the cancellation of $\infty-\infty$ terms 
that are a priori ambiguous, i.e. a priori the first term in the expansion would be $\omega$, not $1/\omega$, but its coefficient is of the 
type $z-z$ and is $k$-independent, therefore disappears under our $\partial/\partial k$.
\footnote{Note also that the result in (\ref{eqn:integrand_freq_x}) contains in the $1/\omega$ piece two subtracted terms linear in $u$ that become log divergent 
at $x\rightarrow \infty$ after an integration in $x$. If one allows for cut-offs $\Lambda_1, \Lambda_2$ for the two subtracted terms such that 
$\Lambda_1/\Lambda_2 \rightarrow c\neq 1$, then we can still obtain an ambiguous result in the final answer (\ref{finalenergy}). Such an ansatz, with 
$\Lambda_1/\Lambda_2=2$ instead of 1 for instance, leading to a difference of $2\ln 2$ in (\ref{finalenergy}), was considered often starting with 
\cite{Gromov:2008fy}, but if we only allow $\Lambda_1-\Lambda_2=$finite, we don't have an ambiguity (more comments on that at the end of this section). 
Observe that in any case this term is linear in $u$, and the approximation we used was for $u\lsim 1/\omega$, hence a term linear in $u$ is really of at least one 
smaller order in $1/\omega$ in our case. Hence even in the case $\Lambda_1/\Lambda_2 \rightarrow c\neq 1$, we can at least claim that we have eliminated not only 
the {\em a priori} ${\cal O}(\omega)$ ambiguity that was implicit in the calculation, but also the ambiguity of the strict $1/\omega$ term (the piece not 
proportional to $u$), and to go beyond that we would need to avoid the constraint $u\lsim 1/\omega$ which we needed solely in order 
to be able to calculate, but was not a theoretical restriction.}

Then we can check that at $x\rightarrow 0$, the coefficient of the $1/\omega$ term becomes regular (constant), 
whereas from $1/\omega^3$ on, we have inverse powers of $x$ at $x\rightarrow 0$, meaning a divergence in the integral $\int_0 dx$. 
Note that these singular terms all come multiplied by powers of $u$, so we cannot properly analyze them using our method, as $u<1/\omega$ for us 
(for technical reasons).

However, we have  
\be
e_{sing}^{sum}(n)=e^{int}_{reg}\left(x=\frac{n}{\omega}\right),
\ee
as expected. 

Similarly, in $e^{int}(x)$ have terms with inverse powers of $x$, which become singular (divergent) when integrated, but we can easily 
verify that
\be
e^{int}_{sing}(x)=e^{sum}_{reg}(n=\omega x).
\ee

Due to this fact, in \cite{McLoughlin:2008he} it was proposed to just drop these singular terms, but this procedure hides a regularization ambiguity, 
since for instance we could expand in a slightly different parameter that $\omega$ and then by the same logic resolve to drop a different divergent 
piece from the total result. With our procedure, it becomes clear that result is unambigous and free of potential divergences, and we are in fact 
led to drop the singular terms of \cite{McLoughlin:2008he}. Indeed, the effect of summing over (\ref{sumsummand}) and integrating over $k$ with 
zero constant is (to leading order in $u$, which is what we can check) the same as just dropping the divergent terms in (\ref{sumprev}).

In conclusion, we see that there were a priori $\infty -\infty$ ambiguities that were hidden in the formal $1/\omega$ expansion procedure above, 
but we have checked that our physical principle just cancels them, and then we can continue with the same calculation as in \cite{McLoughlin:2008he}. 
Namely, the one-loop correction is now 
\be
E^{(1)}=E_{n=0}+\sum_{n\geq 1}e_{reg}^{sum}+\int dx e^{int}_{reg}(x),
\ee
where $E_{n=0}$ is the zero mode contribution. The terms giving odd powers of $J$ are 
\be
E_{n=0}+\int dx e^{int}_{reg}(x)=S+J+\frac{\bar h^2(\bar \lambda)k^2}{2J}u(1+u)+{\cal O}\left(\frac{1}{J^3}\right),\label{finalenergy}
\ee
where 
\be
\bar h(\bar\lambda)=\sqrt{\bar\lambda}-\ln 2+{\cal O}\left(\frac{1}{\sqrt{\bar\lambda}}\right)=2\pi\left(\sqrt{\frac{\lambda}{2}}
-\frac{\ln 2}{2\pi}+{\cal O}\left(\frac{1}{\sqrt{\lambda}}\right)\right)=2\pi h(\lambda),
\ee
agrees with the value of $h(\lambda)$ argued in \cite{McLoughlin:2008he} to be predicted by AdS/CFT
(though a direct calculation of quantum corrections to the dual to $h(\lambda)$ is still lacking).

Note however that changing both the $h(\lambda)$ above and the energy correction simultaneously could maintain agreement
(see e.g. \cite{Bandres:2009kw,Abbott:2010yb}). Here we will assume, 
following \cite{McLoughlin:2008he}, that the choice of $h(\lambda)$ above is unambigous (at least as long as the number of modes summed over 
in various terms differs only by a finite amount; in the heavy-light prescriptions used for instance in \cite{Gromov:2008fy}, 
some terms are summed over twice as many modes than other terms, due to some unitarity prescription).

\section{Conclusions}

In this paper we have proposed to apply the physical principle developed in \cite{Nastase:1998sy} for elimination of ambiguities in the quantum 
corrections to the energy of two dimensional solitons, to the case of classical (long) 
strings moving in gravitational backgrounds, taking as a primer the 
case of the spinning string in $AdS_4\times \mathbb{CP}^3$. In that case, it was found that there existed a certain regularization dependence, 
giving rise to different results (e.g \cite{McLoughlin:2008he} and \cite{McLoughlin:2008ms}). A procedure was devised in \cite{McLoughlin:2008he} that gave a result consistent with AdS/CFT, but the 
regularization issue was hidden, without a clear physical principle to explain the choice. As the long history of the quantum corrections to 
the energy of two dimensional kinks has shown, just because a certain regularization choice seems natural is no guarantee that it is correct, and
one needs some physical input to justify it. 

It was our goal to justify the choice in \cite{McLoughlin:2008he} by a physical principle which can be applied to other cases of long strings 
as well. We have found that technical reasons limit how far we can calculate with our method in this case, but we can check that to leading order in $u$
our procedure eliminates the ambiguities, and therefore justifies the choice in \cite{McLoughlin:2008he}, leading to the result consistent with 
AdS/CFT. We hope to apply the same methods to other long strings in the future.

{\bf Acknowledgements}. We would like to thank Radu Roiban for a careful reading of the manuscript and many useful comments and suggestions.
The work of HN is supported in part by CNPq grant 301219/2010-9. 
CLA would like to thanks Humberto Gomez and Alexis Roa for reading the manuscript and CAPES for full support.

\bibliographystyle{nb}
\bibliography{stringsoliton}
\end{document}